\documentclass[10pt, conference]{IEEEtran}

\usepackage{cite}
\usepackage{amsmath,amssymb,amsfonts}
\usepackage{graphicx}
\usepackage{textcomp}
\usepackage{mathtools}
\usepackage{pgfplots}
\usepackage{textcomp}
\usepackage{tikz}
\pgfplotsset{compat=1.17}

\DeclareMathOperator{\FIT}{FIT}
\DeclareMathOperator{\FITbit}{FIT/bit}

\begin{document}

\IEEEoverridecommandlockouts

% \thanks{
 % 978-1-6654-3274-0/21/\$31.00 \copyright 2021 IEEE. Personal use of this material is permitted.  Permission from IEEE must be obtained for all other uses, in any current or future media, including reprinting/republishing this material for advertising or promotional purposes, creating new collective works, for resale or redistribution to servers or lists, or reuse of any copyrighted component of this work in other works.}

\title{Efficient Error-Correcting-Code Mechanism for High-Throughput Memristive Processing-in-Memory}

\author{
\IEEEauthorblockN{Orian Leitersdorf\IEEEauthorrefmark{1}, Ben Perach, Ronny Ronen, and Shahar Kvatinsky} \IEEEauthorblockA{\emph{Technion - Israel Institute of Technology, Haifa, Israel}} \IEEEauthorblockA{\{orianl\IEEEauthorrefmark{1}, benperach\}@campus.technion.ac.il, ronny.ronen@technion.ac.il, shahar@ee.technion.ac.il}
}

\IEEEaftertitletext{\vspace{-4pt}}

\maketitle

\IEEEpubid{\begin{minipage}{\textwidth}\ \\[12pt] \centering 978-1-6654-3274-0/21/\$31.00 \copyright 2021 IEEE. Personal use of this material is permitted.  Permission from IEEE must be obtained for all other uses, in any current or future media, including reprinting/republishing this material for advertising or promotional purposes, creating new collective works, for resale or redistribution to servers or lists, or reuse of any copyrighted component of this work in other works.
\end{minipage}} 

% ---- Abstract ---- %
\begin{abstract}
Inefficient data transfer between computation and memory inspired emerging processing-in-memory (PIM) technologies. Many PIM solutions enable storage and processing using memristors in a crossbar-array structure, with techniques such as memristor-aided logic (MAGIC) used for computation. This approach provides highly-paralleled logic computation with minimal data movement. However, memristors are vulnerable to soft errors and standard error-correcting-code (ECC) techniques are difficult to implement without moving data outside the memory. We propose a novel technique for efficient ECC implementation along \emph{diagonals} to support reliable computation inside the memory without explicitly reading the data. Our evaluation demonstrates an improvement of over eight orders of magnitude in reliability (mean time to failure) for an increase of about $\mbox{\boldmath $26$}$\% in computation latency. 
\end{abstract}

\begin{IEEEkeywords}
Processing-in-memory (PIM), memristor, memristor aided logic (MAGIC), soft errors, reliability, error-correcting-code (ECC).
\end{IEEEkeywords}

% \vspace{-10pt}

% ---- Introduction ---- %
\section{Introduction}
\label{section:introduction}

\IEEEpubidadjcol

% Inspiration involving bottleneck
Modern computing systems generally involve a separation of computation from memory, as seen in the von Neumann architecture. There have been many efforts at improving the processing and memory units independently, yet recently it seems that the majority of time and energy is spent on the data transfer between them~\cite{DarkMemory}. This presents a performance bottleneck known as the \emph{von Neumann bottleneck} or the \emph{memory wall}, and approaches for resolution include reducing the distance between computation and memory~\cite{IntelligentRAM} or employing several cache memory levels. While these solutions do alleviate the issue, they still require the fundamental need for data transfer between computation and memory. 

% Processing-in-memory solutions, specifically memristor-based ones
\emph{Processing-in-memory} (PIM) is an emerging solution which introduces memory technologies that support both data storage and computation in the same place, potentially eliminating the bottleneck. An attractive implementation of PIM is a resistive memory array, which employs the memristor~\cite{Memristor} as the basic unit of memory and computation. The memristor is a device that represents data via its resistance, either Low Resistive State (LRS) or High Resistive State (HRS), with the unique property where applied voltage changes the resistance. Memristors have several appealing characteristics, including their non-volatility, low power consumption, high speed, and high density in a crossbar array structure~\cite{DesiredMemristor}.

% mMPU and MAGIC
Computation in resistive memory arrays can be based on the concept of stateful logic: representing data with resistance and performing calculations using memristors. One such computation technique is \emph{Memristor-Aided loGIC} (MAGIC)~\cite{MAGIC}, which performs functionally complete logic gates, such as NOR and NOT, inside memristive crossbar arrays. MAGIC operations can be performed in parallel on all the rows or columns of the crossbar array, enabling PIM with massive parallelism.

% Reliability issues
Memristors are vulnerable to soft errors originating from the diffusion of oxygen vacancies (leading to state drift)~\cite{RRAMRefresh}, ion strikes~\cite{IonStrikeLiu, SEISoftErrors}, and environmental factors~\cite{MemristorInfluence}. Memristive PIM solutions are vulnerable since soft errors may change the operands of subsequent computations undetected (leading to incorrect results and wasted time/energy). Memory soft errors are traditionally addressed with error correcting code (ECC), a technique that uses redundant information to detect and correct errors~\cite{SoftErrorTrends, ECCMemristor}. ECC can be implemented along data transfer (computed along write and checked along read) in traditional memories with tolerable overhead~\cite{SoftErrorTrends, ECCMemristor}. However, stateful logic techniques such as MAGIC do not move the data outside of the crossbar array during computation and therefore present a challenge for ECC implementations. Furthermore, performing multiple logic operations in parallel within the memory crossbar array renders ECC difficult to update/check in-memory without hindering efficiency (since large amounts of data can be accessed or changed at once). 

\IEEEpubidadjcol

% Proposed solution and paper overview
We propose a novel implementation of ECC in MAGIC-based crossbar arrays, centered around continuous ECC calculation (update and check) along \emph{diagonals} within the memory. This follows from the fact that MAGIC operations are performed in parallel across rows and columns, leading to diagonals having interesting characteristics. Since additional diagonal wires in the crossbar are not feasible, we mimic their effect through barrel-shifters. We propose extensions to the crossbar array to support this technique and show a significant improvement in reliability (mean time to failure) of over eight orders of magnitude for a modest increase in computation latency of about $26$\%. This paper makes the following contributions: 
\begin{itemize}
    \item Discussion of ECC techniques and their implementation in memristive memory processing units.
    \item Proposal of an efficient in-memory ECC mechanism that continuously updates/checks multi-dimensional parity.
    \item Evaluation of reliability for the proposed ECC solution compared to standard crossbar array, demonstrating over eight orders of magnitude improvement.
\end{itemize}

\vspace{-5pt}

% ---- Background ---- %
\section{Background}
\label{section:background}

\vspace{-3pt}

% MAGIC - Introduction to MAGIC operations, emphasis on parallelism, mMPU
\subsection{Memristor Aided Logic (MAGIC)}
\label{section:background:MAGIC}
\begin{figure}[!t]
\centering 
\includegraphics[width=3.1in]{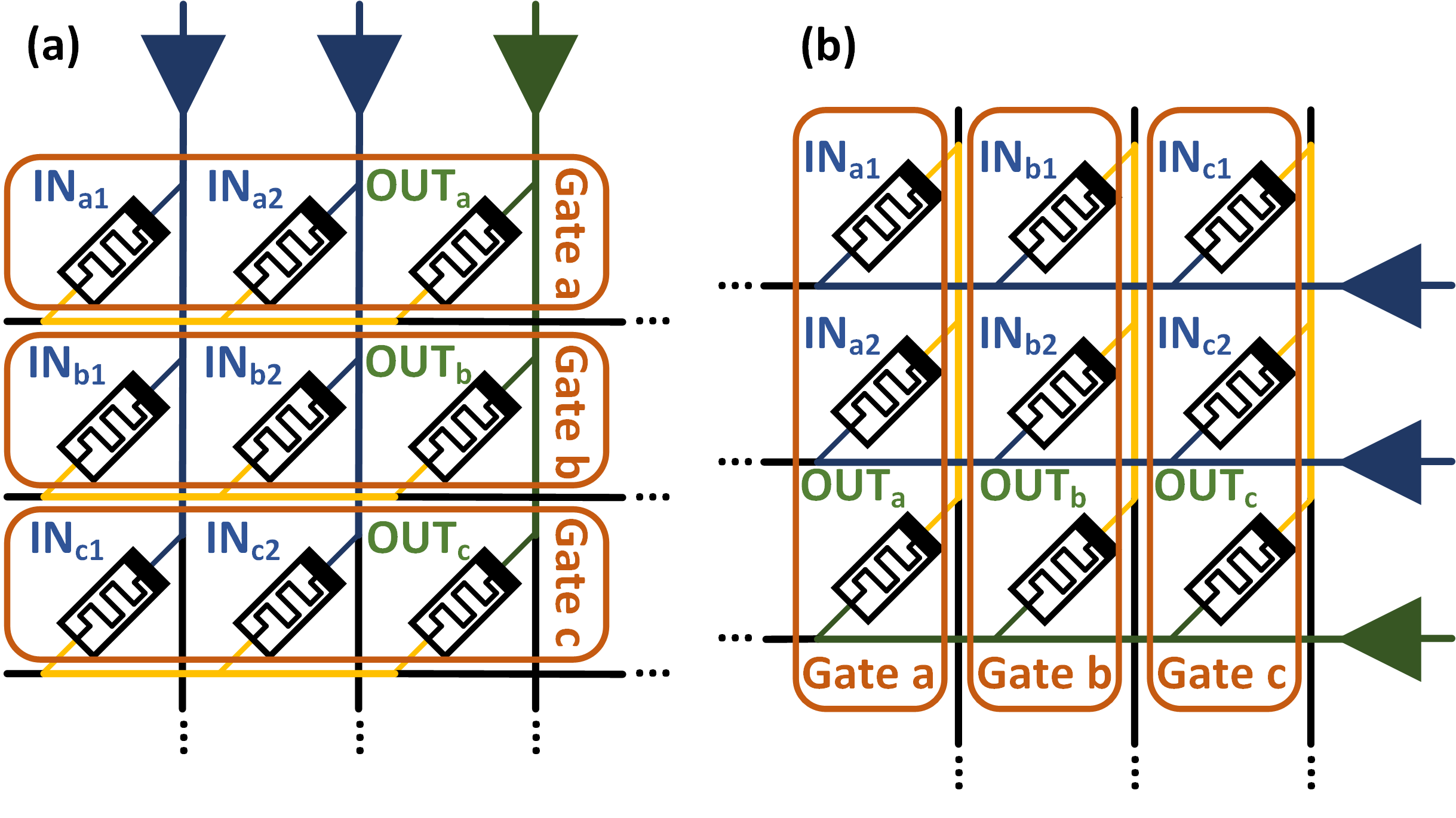}
\caption{Memristive crossbar array with NOR operations in (a) rows and (b) columns. Each scenario performed simultaneously in a single clock cycle.}
\label{fig:crossbar} 
\vspace{-15pt}
\end{figure}

Standard memristor crossbar arrays involve horizontal wordlines, vertical bitlines, and memristors at the crosspoints. Stateful logic techniques for supporting in-memory processing in these crossbars represent data with resistance and perform calculations using memristors. One such example is \emph{Memristor-Aided loGIC} (MAGIC)~\cite{MAGIC}, where logic gates such as NOR/NOT can be calculated between memristors in the same row/column. The gate is performed by controlling the voltages along the wordlines and bitlines in such a way that exploits input/output memristor properties. This solution is especially attractive as it supports \emph{massive parallelism}: the same in-row (in-column) gate can be performed along multiple rows (columns) at the same time with the same applied voltages, as demonstrated in Figure~\ref{fig:crossbar}. This parallelism can be exploited to reduce computation latency~\cite{SIMPLE} and increase throughput for Single Instruction Multiple Data (SIMD) operations~\cite{SIMPLER}.

The overall memory is typically divided into numerous crossbars, connected with CMOS. For example, the memristive Memory Processing Unit (mMPU)~\cite{mMPU} is divided into multiple \emph{banks}, each of which consists of multiple crossbars. This paper is on a per-crossbar-array basis, \textit{i.e.}, the proposed extensions are applied to every crossbar array in the memory.

% Crossbar figure showing MAGIC operations

\vspace{-5pt}

% Memristor soft errors
\subsection{Soft Errors in Memristors}
\label{section:background:memristorSoftErrors}

Similar to other memory technologies~\cite{SoftErrorTrends}, memristors are vulnerable to soft errors (unintentional temporary changes in logical state). Causes for these errors include diffusion of oxygen vacancies (leading to state drift)~\cite{RRAMRefresh}, ion strikes~\cite{IonStrikeLiu, SEISoftErrors}, and environmental factors~\cite{MemristorInfluence}. These soft errors can accumulate over time~\cite{RRAMRefresh, IonStrikeLiu, HfO2PhysicalMechanism} or occur abruptly~\cite{MemristorInfluence, SEISoftErrors}.

Memristor-based PIM is vulnerable to soft errors since the erroneous data affects future in-memory computation undetected, leading to incorrect results.
Previously, a refresh mechanism was proposed to improve the reliability of memristive memory~\cite{RRAMRefresh}. This solution entails periodic full-memory refreshes, periodically resetting the accumulated drift and avoiding such errors. Yet, there are still accumulated soft errors that cannot be addressed with refresh (those which occur between the periodic refreshes). Refresh also does not address abrupt soft errors. Note that refresh can still be used in conjunction with the mechanism proposed in this paper.

\vspace{-10pt}

% ECC Techniques: The trivial horizontal technique, and an introduction into the diagonal technique.
\section{Proposed ECC Technique}
\label{section:proposedTechnique}
Error correcting code (ECC) employs redundant data (check-bits) to improve reliability and is used in memories to avoid silent data corruption~\cite{SoftErrorTrends, ECCMemristor}. ECC may provide PIM memory with the ability to detect/correct soft errors.

We assume a PIM model where a desired function is performed by the memory controller translating the function to a sequence of in-memory logic gates executed by the memristive memory cells~\cite{mMPU}. Implementation of such functions involves input, intermediate, and output memristors (where the intermediates store temporary data used only within the function). We focus on soft errors that occur in the function input memristors before the function is performed and updating the ECC afterward to reflect the newly-stored function output data. Soft errors in intermediate memristors and incorrect MAGIC gate calculations are left for future work. 

In non-processing memories, ECC can be implemented along data transfer using CMOS logic (with low cost compared to data transfer)~\cite{SoftErrorTrends, ECCMemristor}. That is, the ECC can be computed when data is written and checked when read. Yet, PIM memories make such an implementation not possible as data can be used and altered within the memory (without being explicitly read). Furthermore, stateful-logic parallelism makes many implementations impractical as large amounts of data can be used and altered simultaneously. Instead, we seek an ECC mechanism which efficiently supports in-memory logic operations by using the same principles as the computation, \textit{i.e.}, by using paralleled memristor-based computation rather than external CMOS circuit. This implementation must be capable of both continuously updating the ECC as MAGIC operations change the data, and also detecting/correcting errors before they are accessed as inputs for a MAGIC gate.

% Figure demonstrating both the horizontal and the diagonal solutions
 \begin{figure*}[!t]
 \centering 
 \includegraphics[width=6.1in]{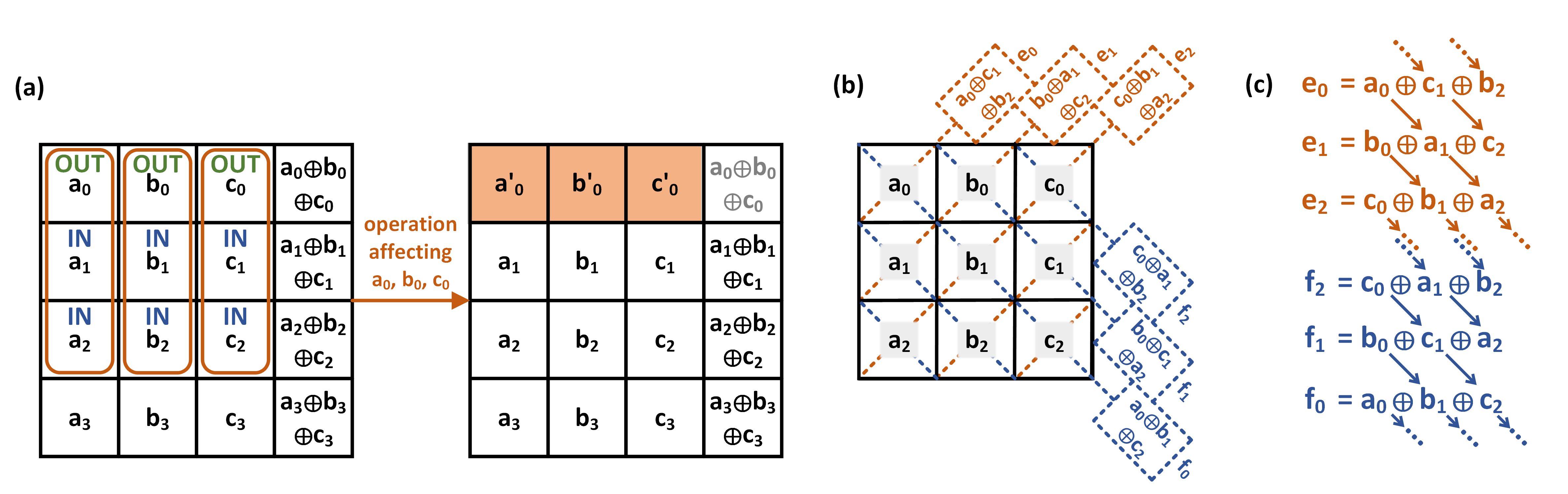} 
 \caption{(a) ECC for PIM when check-bits are calculated horizontally. The top-right check-bit had all $n$ of its data-bits updated in the single-cycle logic operation (which included $n=3$ gates), requiring $\Theta(n)$ cycles to recalculate the check-bit. (b) Proposed solution where check-bits are calculated on both leading (orange) and counter (blue) diagonals ($n=3$). (c) The shift pattern that emerges in the diagonal ECC: the letters shift by column index.}
 \label{fig:proposedSolutions}
 \vspace{-10pt}
 \end{figure*}

A possible solution to support ECC for PIM would be to calculate check-bits horizontally. For example, dividing the memory into horizontal bytes and declaring the eighth bit of every byte as a parity bit. Figure~\ref{fig:proposedSolutions}(a) illustrates this approach for 4 bits rather than a byte. When performing a single MAGIC operation in a row, across $n$ rows (similar to Figure~\ref{fig:crossbar}(a)), we find that the check-bits could be updated using $\Theta(1)$ operations since for any check-bit only up to one of the representing data-bits was changed. Yet, when performing an operation in a column, across $n$ columns (similar to Figure~\ref{fig:crossbar}(b)), we find that $\Theta(n)$ operations would be required to update the check-bits -- significantly hindering efficiency. Here a single check-bit had all $n$ of its data-bits altered in this single operation.

Instead, we propose a unique technique of calculating check-bits along wrap-around \emph{diagonals}. This solution guarantees that for any check-bit, only up to one data-bit can be altered in any paralleled operation. This possibly remedies the problem with the previous solution and suggests that an $\Theta(1)$ solution is possible. In addition, we find that this solution can be calculated independently on \emph{both} \emph{leading} (bottom left to top right) and \emph{counter} (bottom right to top left) diagonals, to provide error-correction capabilities through multi-dimensional codes~\cite{MultidimensionalParity}. We can visualize these placements by adding check-bits on diagonals as illustrated in Figure~\ref{fig:proposedSolutions}(b), yet in Section~\ref{section:architecture} we show that the check-bits are placed in a dedicated extension. This technique trades off reliability, complexity, and overhead in the following considerations:
\begin{itemize}
\item \textit{The code used for check-bits along a diagonal of data-bits}. Increased complexity leads to increased reliability at the cost of more complex calculations and more overhead. 
\item \textit{The number of data-bits in each diagonal code}. We divide the $n \times n$ crossbar in an imaginary grid of $m \times m$ $\emph{blocks}$, with check-bits for every diagonal (leading and counter) of data-bits in every block. Smaller blocks increase overall reliability at the cost of more data overhead. 
\end{itemize}

We choose to implement a \emph{parity code} for all leading and counter diagonals on all blocks. This provides single-error-correction capability~\cite{MultidimensionalParity} on a per-block basis since any soft error leaves a signature of the affected leading/counter diagonals -- which can uniquely identify the single error\footnote{Note $m$ must be odd for $m \times m$ blocks to have wrap-around diagonals uniquely index cells. Otherwise, two diagonals may intersect in two locations.}. While this diagonal technique inherently supports updating the check-bits after a MAGIC operation, verifying the check-bits (to correct errors) requires a separate process. We propose specific ECC-checking before logic function execution, to verify that the inputs are correct, and periodic full-memory checks to address soft errors that occur in areas that are rarely accessed. Periodic checks \emph{alone} cannot solve the problem due to the possibility of accessing data between checks.

This Section considered the concept of continuous parity: check-bits are updated when only some of the data changes, using only the difference between the old/new data values without the other data-bits. Conversely, ECC in traditional memories is typically computed using all of the relevant data-bits. The partial-update method can lead to a rare scenario where a perfectly correct bit is considered incorrect (false positive). This occurs when writing over a bit that encountered a soft error, before the bit was checked (in either specific/periodic checks). Techniques such as \emph{locally decodable codes}\footnote{https://en.wikipedia.org/wiki/locally\_decodable\_code} can solve this problem and will be investigated in future work. This paper does not deal with this rare case.

\vspace{-10pt}

% ---- Architecture ---- %
\section{Architecture Design}
\label{section:architecture}

% Transition into architecture proposal
We present extensions to a MAGIC crossbar array to incorporate the proposed \emph{diagonal} technique in an efficient manner which supports MAGIC parallelism. Previously the check-bits were visualized as continuations of diagonals, yet additional wires along diagonals substantially complicate the crossbar structure (since memristors have two terminals). Instead, we decide to add an additional memory named Check Memory (CMEM), to \emph{extend} the original crossbar-array memory (MEM). Each bit in the CMEM is a check-bit for some diagonal of a block in the MEM.

The assumed PIM model enables in-memory logical function computation using input memristors, intermediate memristors, and output memristors (see Section \ref{section:proposedTechnique}). When performing a MAGIC operation that writes to the output memristors (writing data that needs to be covered in the ECC), the CMEM needs to be updated as well. For every such operation, called a \emph{critical} operation, the following steps occur:
\begin{enumerate}
    \item \emph{Cancel} the effect of the old data-bits on the check-bits.
    \item \emph{Perform} the critical MAGIC operation in the MEM.
    \item \emph{Add} the effect of the new data-bits on the check-bits.
\end{enumerate}

% Shifter recap and overall structure
The parity code allows for \emph{cancelling} and \emph{adding} effects as desired through an exclusive or (XOR) operation. To perform steps $1$ and $3$, we need to support the transfer of old/new data from the MEM to the CMEM, where an XOR operation between the old/new data and the previous check-bit is performed (and stored as the new check-bit). The CMEM requires that the data arrive according to the diagonal indices, but the MEM's interface (wordlines/bitlines) does not provide this. To accomplish this, we use barrel shifters to emulate the diagonal effect following from the pattern seen in Figure~\ref{fig:proposedSolutions}(c). 

In order to detect/correct errors, we check the ECC on function inputs before function execution. We assume that function inputs are generally concentrated in the same blocks (similar to how data is typically stored in bytes), and we check only those blocks. Since the function can be performed in parallel along rows (columns) for SIMD operation, then we propose a method for checking an entire row (column) of blocks. The row (column) of blocks is copied into the CMEM row-by-row (column-by-column) in $m$ MAGIC NOT operations. Then, while the MEM is free to perform other non-critical operations, the CMEM continues by calculating the XOR of the copied rows (columns). The vector difference between the computed parity and the stored parity, also known as the \emph{syndrome}~\cite{Shooman}, is computed using XOR. Finally, logic in the CMEM determines if the syndrome is non-zero (meaning that a soft error has occurred) and then acts accordingly, with the CMEM controller, to correct the memory soft error.

\begin{figure}[!t]
\centering 
\includegraphics[width=3.3in]{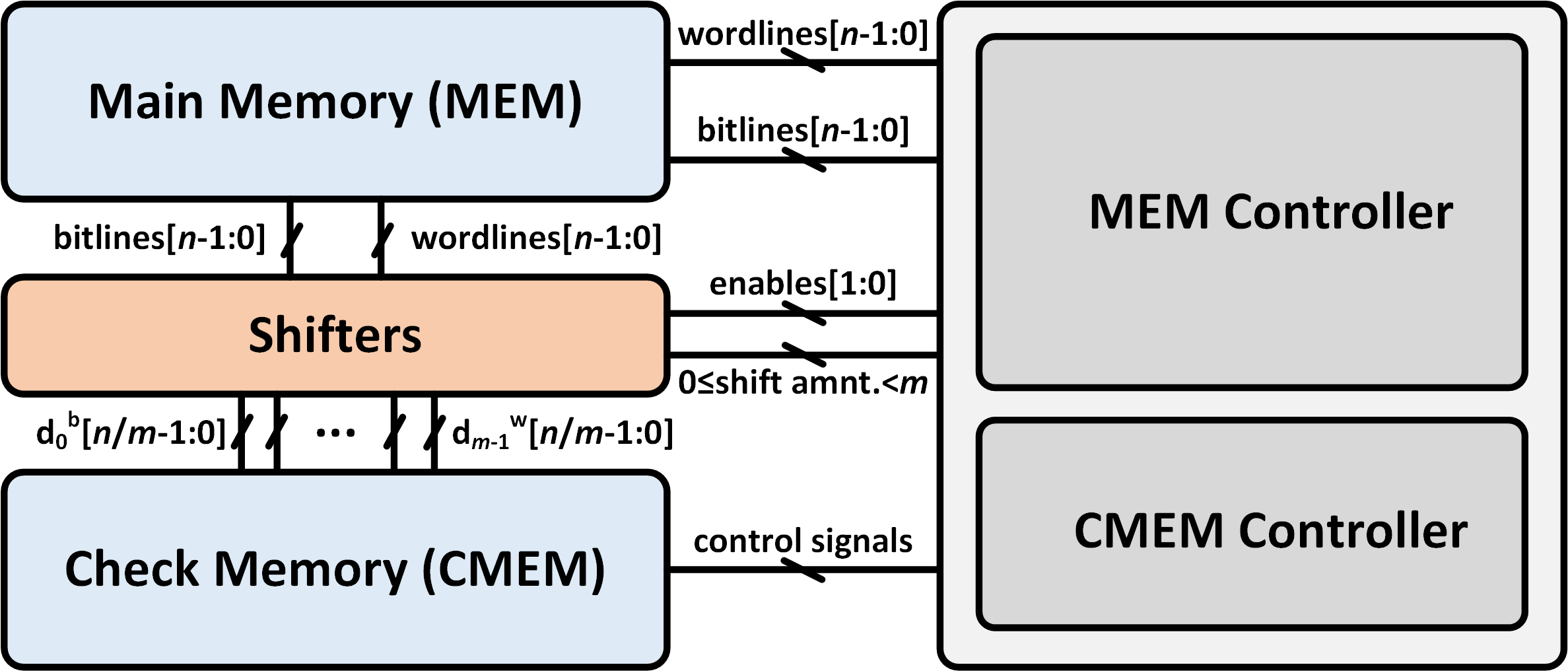}
\caption{Proposed architecture consists of the MEM, CMEM, the connecting shifters, and the controllers. Control signals are as presented in Fig. 4.}
\label{fig:proposedStructure} 
\vspace{-5pt}
\end{figure}

Figure~\ref{fig:proposedStructure} shows the proposed design structure. We now describe in detail the design of each component. Without loss of generality, we focus on leading diagonals.

% Proposed structure figure

\vspace{0pt}

% Explanation that CMEM cannot be implemented using single crossbar
\subsection{Memory Structure}
\label{section:architecture:memoryStructure}

We extend the MEM, a single crossbar array, with the CMEM which stores ECC check-bits for the MEM. While it seems that the CMEM can be implemented via a single crossbar array, the possibility for both in-row and in-column MAGIC operations in the MEM forces the CMEM to be divided into $m$ crossbar arrays according to block diagonal indices. Furthermore, the complexity of XOR execution via NOR operations introduces the desire for dedicated processing crossbars. Thus, we propose the following CMEM division, as illustrated in Figure~\ref{fig:cmem}, that enables pipelining the operations. 

% Explanation of check-bit crossbar format (m crossbars, each for a different diagonal index)
\subsubsection{Check-bit Crossbars}
\label{section:architecture:CMEM:crossbars}
\begin{figure}[!t]
\centering 
\includegraphics[width=3.2in]{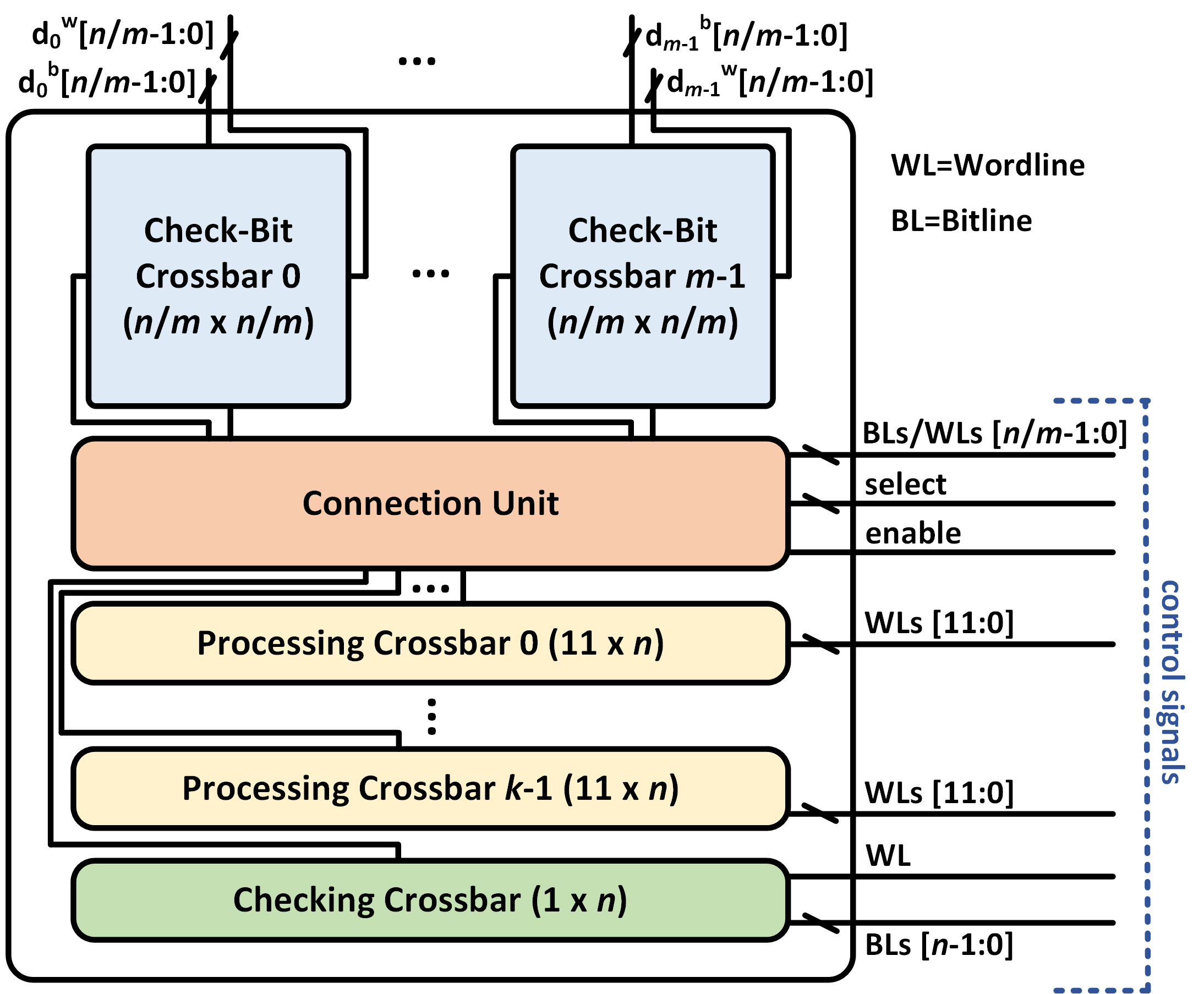}
\caption{Structure of the Check Memory (CMEM): check-bit crossbar arrays (blue), processing crossbars (yellow), checking crossbar (green), and a connection unit (orange). Control signals connect the CMEM to the controller.}
\label{fig:cmem} 
\vspace{-5pt}
\end{figure}

The check-bits are stored in $m$ crossbar arrays labeled $0$ to $m-1$ (Figure~\ref{fig:cmem}). All crossbars are of dimension $(n/m) \times (n/m)$ with the $i^{th}$ crossbar containing the check-bits for the $i^{th}$ diagonal in all blocks. That is, the memristive memory cell $(a,b)$ in check-bit crossbar $i$ stores the check-bit for the $i^{th}$ diagonal of the block which is $a$ blocks from the left and $b$ blocks from the top.

% Discuss how XOR3 is used in the memory.
\subsubsection{XOR3}
\label{section:architecture:CMEM:XOR3}
XOR3 is the main operation supported in the CMEM for both ECC update and ECC check. For ECC update, the CMEM receives from the MEM both the old and the new data-bits, and updates the check-bits to (current-check-bits $\oplus$ old-data-bits $\oplus$ new-data-bits). For ECC checking, the CMEM receives $m$ rows (columns) and computes their XOR using a paralleled XOR3 tree. Note that XOR3 is performed with $8$ MAGIC NOR operations.

% CMEM Figure

% Shifters figure
\begin{figure*}[!th]
\centering 
\includegraphics[width=6in]{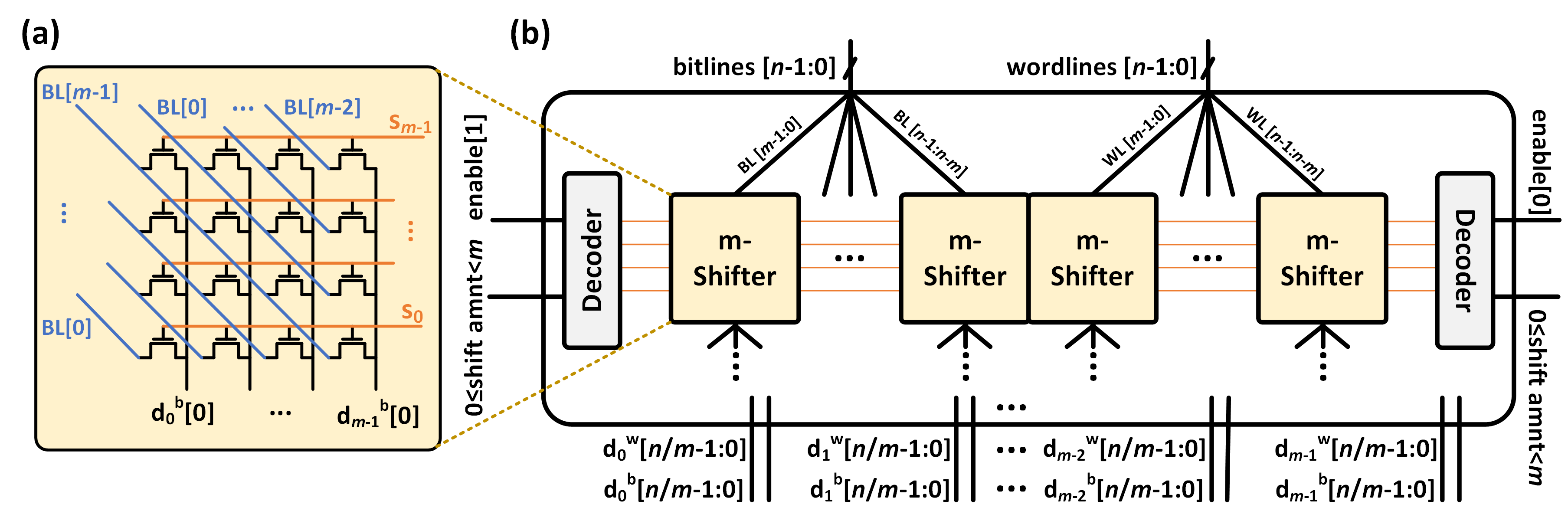}
\caption{(a) The basic implementation unit: $m$-Shifter constructed from transistors, similar to those in~\cite{NNPIM, UltraEfficient}. (b) The combined shifter structure. The inputs are data received from the MEM (either along the bitlines or the wordlines) divided into vectors of size $m$ (\textit{i.e.}, bitline[$m-1$:0], ..., wordline[$n-1$:$n-m$]) corresponding to the data from each block. Each one of these groups can be shifted up to $m$ places, to account for the diagonal effect on a per-block basis.}
\label{fig:shifters} 
\vspace{-7pt}
\end{figure*}

% Discuss pipe-lining of the XOR3 calculation process
\subsubsection{Processing Crossbar}
\label{section:architecture:CMEM:PAs}
Although the trivial implementation is to perform the XOR3 gates in the CMEM crossbars using MAGIC, this effectively stalls the CMEM for $8$ cycles per MEM critical operation and even more for ECC checking. Instead, we present a pipelined model involving separated crossbar arrays named \emph{processing crossbars} which calculate the XOR3 through MAGIC NOR. The old and new data-bits are transferred (with MAGIC NOT) directly from the MEM to a processing crossbar, the old check-bits are transferred from the CMEM crossbars to the same processing crossbar, and then the processing crossbar continues with the calculation for $8$ cycles within the processing crossbar, while the MEM and the CMEM are free to perform other operations\footnote{There are several special corner cases that should be treated. For example, when resetting an entire block then the block's ECC can also be reset directly rather than being calculated through XOR. With regards to subsequent updates in the same block, these can be both avoided in function design and addressed using processing crossbar forwarding.}.

\subsubsection{Checking Crossbar}
\label{section:architecture:CA}
The Checking Crossbar array is a row of memristors responsible for evaluating the syndromes to detect/correct errors. After the syndromes for a row (column) of blocks is computed via the processing crossbars, it is transferred to the checking crossbar. Then, through MAGIC NOR operations \cite{MAGIC}, the syndrome of each block is compared to zero (to see if there are any errors to detect/correct). For cases in which the syndrome is non-zero, sensing circuitry in the CMEM controller (connected directly to the checking crossbar) reads the block syndrome of length $2m$. Then, a logical function is performed in the CMEM controller and the error is detected or corrected (with the correct value being written into the crossbar).

% Discussion of connection unit and crossbar lines
\subsubsection{Connection Unit}
\label{section:architecture:CMEM:connectionUnit}
The connection unit connects between the shifters, the check-bit crossbars, the processing crossbars, and the checking crossbar. The connection unit allows the CMEM controller to apply the necessary voltages on check-bit crossbars by connecting the crossbar lines vector to the bitlines/wordlines of the crossbar arrays. This is possible due to operation symmetry with different check-bit crossbars and this reduces the peripheral circuitry required in the CMEM controller. Implementation of the connection unit is similar to shifters (see Section \ref{section:architecture:shifters}).

% Specific definition of shifter role, accompanying figure showing implementation in full
\subsection{Shifter Structure}
\label{section:architecture:shifters}

The shifters receive the wordlines/bitlines from the MEM and the row (column) index modular $m$. The shifters output $2m$ vectors ${d}_0^w, ..., {d}_{m-1}^w, {d}_0^b, ..., {d}_{m-1}^b$, each of length $n/m$. Each vector ${d}_i^w$ (${d}_i^b$) consists of the data-bits collected along a wordline (bitline) of blocks for diagonals of the $i^{th}$ index. These blocks are chosen according to the row (column) of blocks which are affected by the critical MEM operation.

We construct the shifters using transistors and CMOS decoders (for the modular shift input), as shown in Figure~\ref{fig:shifters}. The shifters merely \emph{reroute} the paths connecting the MEM to the CMEM, meaning that the data transfer between the MEM and the CMEM is similar to a transfer of data within a single crossbar array (thus parallelism and efficiency are kept). Similar shifters have been used in other works~\cite{NNPIM, UltraEfficient}. 

\vspace{-3pt}

% Discuss how the MEM controller is similar to a standard one but with CMOS communication to CMEM
% Discuss how the CMEM controller 1) contains the PA controllers, 2) coordinates with MEM the data transfer.
\subsection{Controllers}
\label{section:architecture:controllers}
Memory controllers in PIM designs extend the traditional interface of reading/writing to include in-memory computation. Controllers for MAGIC-based crossbars, such as the CMOS mMPU controller~\cite{mMPU}, implement this interface by applying voltages on the crossbar array wordlines/bitlines. The MEM controller is similar to these controllers, with the addition of CMOS logic signals for coordinating operations with the CMEM controller. The CMEM controller is also similar to standard controllers since it has indirect access to the check-bit crossbars through the connection unit. Furthermore, the CMEM controller contains the Processing Crossbar (PC) controllers which consist of simple finite state machines that perform the pre-defined XOR3 steps.

\vspace{-2pt}

% ---- Results ---- %
\section{Results}
\label{section:results}

% Comparison of Reliability
\subsection{Reliability}
\label{section:results:reliability}

\begin{figure}[!t]
\centering 
\begin{tikzpicture}
\begin{loglogaxis}[
    xlabel={Memristor Soft Error Rate ($\FITbit$)},
    ylabel={Memory MTTF (hours)},
    xtick={1e-5, 1e-3,1e-1,1e1,1e3},
    ytick={1e18, 1e15, 1e12, 1e9, 1e6, 1e3, 1e0},
    ymin=10^0,
    ymax=10^15,
    legend pos=north east,
    every axis plot/.append style={ultra thick},
    width=\linewidth,
    height=2.3in
]

\addplot[
    color=blue,
    ]
    coordinates {
(0.00001, 12512.)(0.0000372759, 3365.38)(0.00013895, 911.66)(0.000517947, 253.536)(0.0019307, 77.4831)(0.00719686, 32.0482)(0.026827, 24.1399)(0.1, 24.)(0.372759, 24.)(1.3895, 24.)(5.17947, 24.)(19.307, 24.)(71.9686, 24.)(268.27, 24.)(1000., 24.)
    };

\addplot[
    color=brown,
    ]
    coordinates {
(0.00001, 4.68684*10^14)(0.0000372759, 3.37305*10^13)(0.00013895, 2.42754*10^12)(0.000517947, 1.74706*10^11)(0.0019307,  1.25734*10^10)(0.00719686, 9.04888*10^8)(0.026827, 6.51235*10^7)(0.1, 4.68686*10^6)(0.372759, 337318.)(1.3895, 24287.5)(5.17947, 1759.12)(19.307, 138.124)(71.9686, 25.8214)(268.27, 24.)(1000., 24.)
    };
\addplot[color=black, dashed] coordinates {(10^(-3), 10^0) (10^(-3), 10^15)};
  
\legend{Baseline, Proposed ECC}
\end{loglogaxis}
\end{tikzpicture}
\caption{1GB Memory Mean-Time-To-Failure (MTTF) sensitivity analysis of proposed and control designs for varying memristor Soft Error Rates (SERs). For reference, Flash memory SER is approximately $10^{-3}\cdot\FITbit$ \cite{SoftErrorTrends}.}
\label{fig:reliabilityResults}
\vspace{-12pt}
\end{figure}
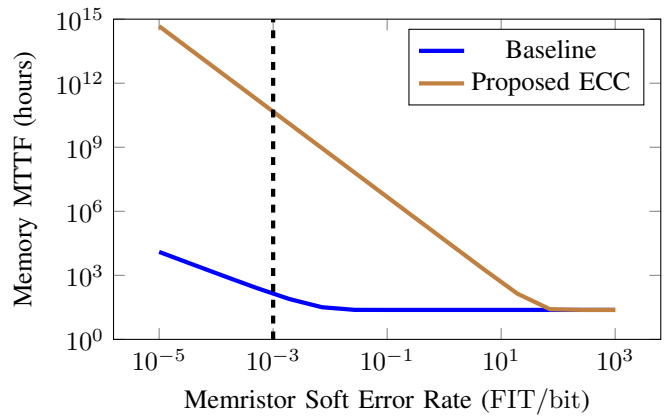

We seek to evaluate the reliability improvements of the proposed architecture compared to a baseline of no ECC. In the proposed model, ECC is implemented on a \emph{per-block} basis that is capable of single-error-correction~\cite{MultidimensionalParity}. ECC is checked both periodically and on function inputs. For the reliability analysis, we assume the worst case in which the time between ECC checks for a specific bit is this ECC-checking period.

\vspace{0pt}

We define the sensitivity analysis as follows. Assume soft errors of memristors are distributed uniformly and independently and have a constant Soft Error Rate (SER) $\lambda~[\FITbit]$ (where $1\cdot\FITbit$ represents one error per $10^9$ hours in a specific memristor). We assume that the full-memory checks are performed every $T=24$ hours (chosen to have negligible performance impact while still providing adequate reliability), and find that the probability that a specific memristor encounters a soft error in $T$ hours is $1-e^{-\lambda T / 10^{9}}$~\cite{Shooman}. We assume the memory is sized $n \times n$ for $n=1020$, and that blocks are sized $m \times m$ for $m=15$.

On a per-block basis, we find that there will be no uncorrected errors in a specific block (block \emph{success}) if there are either zero or one memristor soft errors. This probability of block success is computed using the binomial distribution. Since the blocks are independent, then we find the probability that the $n \times n$ memory succeeds to be the probability of single block success raised to the power of the number of blocks. For this analysis, we consider multiple $n\times n$ crossbars connected to form a 1GB memory. We find the 1GB memory reliability as the $n \times n$ crossbar reliability raised to the power of the number of crossbars. Then, we compute the memory failure rate [$\FIT$] as the probability of memory failure in $T$ hours multiplied by $10^9 / T$ (finding the number of memory failures per $10^9$ hours). Figure~\ref{fig:reliabilityResults} shows the memory Mean-Time-To-Failure (MTTF) as determined from the memory failure rate ($10^9$ divided by the failure rate~\cite{Shooman}). For a memristor SER of $10^{-3}\cdot\FITbit$ (similar to Flash memory~\cite{SoftErrorTrends}), we find an improvement in MTTF by a factor of over $3 \cdot 10^8$.

\vspace{-5pt}

% Benchmarks for cycle count, shows minor difference
\subsection{Latency}
\label{section:results:time}
To evaluate the latency overhead of the proposed mechanism, we generated different logic functions using the SIMPLER tool~\cite{SIMPLER}. SIMPLER constructs a sequence of MAGIC NOR operations to perform any logic function in the memory within a single row. An extension of the SIMPLER algorithm is developed to consider the additional operations required in the proposed architecture (checking ECC on inputs and updating ECC for the outputs)\footnote{Code available at https://github.com/oleitersdorf/ecc-simpler-magic.}.
The adapted tool first runs the SIMPLER algorithm and then schedules the operations needed in the proposed method through a greedy algorithm that checks MEM/CMEM availability (adding cycles if they are not available when an operation needs to occur). Table~\ref{table:timeBenchmarks} lists the baseline versus the proposed technique in terms of the number of cycles for the EPFL benchmarks\cite{EPFL}.

\begin{table}[!t]
\caption{Latency (Clock Cycles)}
\centering
\begin{tabular}{l|c|c|c|cc}
 Benchmark&Baseline & Proposed & Overhead (\%)& PC (\#)\\
 \hline
 adder&1531&2050&34.0&3\\
 arbiter&12798&13316&4.05&2\\
 bar&4051&4510&11.3&4\\
 cavlc&841&879&4.5&3\\
 ctrl&134&201&50.0&5\\
 dec&360&1101&205.8&8\\
 int2float&295&324&9.83&3\\
 max&4200&5101&21.5&4\\
 priority&730&876&20.0&3\\
 sin&7919&7995&0.96&3\\
 voter&12738&13733&7.81&2\\
 \hline
 Geo. Mean&&&26.23&3.36\\
\end{tabular}
\label{table:timeBenchmarks}
\vspace{-10pt}
\end{table}

\begin{table}[!t]
\centering
\caption{Memristor/Transistor Count}
\begin{tabular}{l|c|c|c}
 Unit & \# Memristor & \# Transistor & Expression\\
 \hline
 Data (MEM) & $1.04 \cdot 10^{6}$ & 0 & $n \times n$\\
 Check-Bits & $1.39 \cdot 10^{5}$ & 0 & $2 \times m \times (n/m)^2$\\
 Processing XBs & $6.73 \cdot 10^4$ & 0 & $2 \times 11 \times k \times n$\\
 Checking XB & $2.04 \cdot 10^3$ & 0 & $2 \times n$\\
 Shifters & 0 & $6.12 \cdot 10^4$ & $4 \times n \times m$\\
 Connection Unit & 0 & $1.43 \cdot 10^4$ & $2 \times n \times (k+4)$\\
 \hline
 Total&$1.25 \cdot 10^{6}$&$7.55 \cdot 10^{4}$
\end{tabular}
\label{table:memristorTransistorCount}
\vspace{-10pt}
\end{table}

Our results show that the overhead to support ECC is relatively low, with the exception of the decoder benchmark (since it involves a dense and long sequence of critical operations). We also present the minimal number of processing crossbars (PCs) required to perform the benchmark, noting that we need at most eight processing crossbars (to support any logic function without stalling due to lack of processing crossbars).

\vspace{-5pt}

% Memristor and transistor counts, area outside the scope
\subsection{Area}
To evaluate the additional area required for our proposed design, we assess memristor and transistor counts for a case study of $n=1020$, $m=15$ with $k=3$ processing crossbars. Table~\ref{table:memristorTransistorCount} details the distribution of the memristors and transistors counts in the proposed architecture, focusing on the MEM, shifters, and the CMEM. These device counts constitute a preliminary analysis into the additional area requirements, with specific layout and area analysis left for future work.

% ---- Conclusion ---- %
\vspace{-3pt}
\section{Conclusion}
\label{section:conclusion}
We present a novel in-memory ECC mechanism that improves reliability substantially. The architecture is based on a unique technique of determining ECC along \emph{diagonals}, continuously updating and checking, to support stateful-logic parallelism. We demonstrate a significant improvement in mean time to failure by eight orders of magnitude for a modest increase of approximately $26\%$ in latency. Full layout and circuit design are left for future work.

\vspace{-3pt}

\section*{Acknowledgment}
This work was supported in part by the European Research Council through the European Union's Horizon 2020 Research and Innovation Programe under Grant 757259, and in part by the Israel Science Foundation under Grant 1514/17.

\vspace{-3pt}

% ---- References ---- %
\bibliographystyle{IEEEtran}
\bibliography{main}

% Generated by IEEEtran.bst, version: 1.14 (2015/08/26)
\begin{thebibliography}{10}
\providecommand{\url}[1]{#1}
\csname url@samestyle\endcsname
\providecommand{\newblock}{\relax}
\providecommand{\bibinfo}[2]{#2}
\providecommand{\BIBentrySTDinterwordspacing}{\spaceskip=0pt\relax}
\providecommand{\BIBentryALTinterwordstretchfactor}{4}
\providecommand{\BIBentryALTinterwordspacing}{\spaceskip=\fontdimen2\font plus
\BIBentryALTinterwordstretchfactor\fontdimen3\font minus
  \fontdimen4\font\relax}
\providecommand{\BIBforeignlanguage}[2]{{%
\expandafter\ifx\csname l@#1\endcsname\relax
\typeout{** WARNING: IEEEtran.bst: No hyphenation pattern has been}%
\typeout{** loaded for the language `#1'. Using the pattern for}%
\typeout{** the default language instead.}%
\else
\language=\csname l@#1\endcsname
\fi
#2}}
\providecommand{\BIBdecl}{\relax}
\BIBdecl

\bibitem{DarkMemory}
A.~{Pedram} \emph{et~al.}, ``Dark memory and accelerator-rich system
  optimization in the dark silicon era,'' \emph{IEEE Design \& Test}, vol.~34,
  no.~2, 2017.

\bibitem{IntelligentRAM}
D.~Patterson \emph{et~al.}, ``A case for intelligent {RAM},'' \emph{Micro,
  IEEE}, vol.~17, pp. 34 -- 44, 04 1997.

\bibitem{Memristor}
L.~{Chua}, ``Memristor-the missing circuit element,'' \emph{IEEE Transactions
  on Circuit Theory}, vol.~18, no.~5, pp. 507--519, 1971.

\bibitem{DesiredMemristor}
S.~{Kvatinsky}, E.~G. {Friedman} \emph{et~al.}, ``The desired memristor for
  circuit designers,'' \emph{IEEE CAS}, vol.~13, no.~2, pp. 17--22, 2013.

\bibitem{MAGIC}
S.~{Kvatinsky} \emph{et~al.}, ``{MAGIC}—memristor-aided logic,'' \emph{IEEE
  Trans. Circuits Syst., II, Exp. Briefs}, vol.~61, no.~11, pp. 895--899, 2014.

\bibitem{RRAMRefresh}
A.~M.~S. {Tosson} \emph{et~al.}, ``{RRAM} refresh circuit: A proposed solution
  to resolve the soft-error failures for {HfO2/Hf} {1T1R} {RRAM} memory cell,''
  in \emph{GLSVLSI}, 2016, pp. 227--232.

\bibitem{IonStrikeLiu}
R.~{Liu} \emph{et~al.}, ``Investigation of single-bit and multiple-bit upsets
  in oxide {RRAM}-based {1T1R} and crossbar memory arrays,'' \emph{IEEE
  Transactions on Nuclear Science}, vol.~62, no.~5, pp. 2294--2301, 2015.

\bibitem{SEISoftErrors}
D.~{Mahalanabis} \emph{et~al.}, ``Investigation of single event induced soft
  errors in programmable metallization cell memory,'' \emph{IEEE Transactions
  on Nuclear Science}, vol.~61, no.~6, pp. 3557--3563, 2014.

\bibitem{MemristorInfluence}
N.~Wald and S.~Kvatinsky, ``Understanding the influence of device, circuit and
  environmental variations on real processing in memristive memory using
  memristor aided logic,'' \emph{MEJ}, vol.~86, 02 2019.

\bibitem{SoftErrorTrends}
C.~{Slayman}, ``Soft error trends and mitigation techniques in memory
  devices,'' in \emph{RAMS}, 2011, pp. 1--5.

\bibitem{ECCMemristor}
D.~{Niu} \emph{et~al.}, ``Low power memristor-based {ReRAM} design with error
  correcting code,'' in \emph{ASP-DAC}, 2012, pp. 79--84.

\bibitem{SIMPLE}
R.~{Ben Hur} \emph{et~al.}, ``{SIMPLE MAGIC}: Synthesis and in-memory mapping
  of logic execution for memristor-aided logic,'' in \emph{ICCAD}, 2017.

\bibitem{SIMPLER}
R.~{Ben-Hur} \emph{et~al.}, ``{SIMPLER MAGIC}: Synthesis and mapping of
  in-memory logic executed in a single row to improve throughput,'' \emph{IEEE
  TCAD}, vol.~39, no.~10, pp. 2434--2447, 2020.

\bibitem{mMPU}
N.~Talati \emph{et~al.}, \emph{mMPU---A Real Processing-in-Memory Architecture
  to Combat the von Neumann Bottleneck}.\hskip 1em plus 0.5em minus 0.4em\relax
  Springer, 2020, pp. 191--213.

\bibitem{HfO2PhysicalMechanism}
H.~{Chang} \emph{et~al.}, ``Physical mechanism of {HfO2}-based bipolar
  resistive random access memory,'' in \emph{VTSA}, 2011, pp. 1--2.

\bibitem{MultidimensionalParity}
J.~M. Shea and T.~F. Wong, \emph{Multidimensional Codes}.\hskip 1em plus 0.5em
  minus 0.4em\relax John Wiley \& Sons, Inc., 2003.

\bibitem{Shooman}
M.~L. Shooman, \emph{Reliability of Computer Systems and Networks: Fault
  Tolerance, Analysis, and Design}.\hskip 1em plus 0.5em minus 0.4em\relax USA:
  John Wiley \& Sons, Inc., 2002.

\bibitem{NNPIM}
S.~{Gupta} \emph{et~al.}, ``{NNPIM}: A processing in-memory architecture for
  neural network acceleration,'' \emph{IEEE TC}, vol.~68, no.~9, 2019.

\bibitem{UltraEfficient}
M.~{Imani} \emph{et~al.}, ``Ultra-efficient processing in-memory for data
  intensive applications,'' in \emph{DAC}, 2017, pp. 1--6.

\bibitem{EPFL}
L.~Amarù \emph{et~al.}, ``The {EPFL} combinational benchmark suite,''
  \emph{IWLS}, 2015.

\end{thebibliography}

\end{document}